\documentclass[aps,prl,twocolumn,showpacs,superscriptaddress,amsmath,amssymb,10pt]{revtex4-1}

\usepackage{graphicx}
\DeclareGraphicsExtensions{.pdf,.png,.jpg}

\usepackage[normalem]{ulem}
\usepackage{xcolor}

\begin{document}

\title{\Large{Longitudinal instabilities affecting the moving critical layer laser-plasma ion accelerators}}

\author{Aakash A. Sahai}
\email{aakash.sahai@gmail.com}
\author{Thomas C. Katsouleas}
\affiliation{Dept of Electrical Engineering, Duke University, Durham, NC, 27708 USA}

\begin{abstract}
In this work we analyze the longitudinal instabilities of propagating acceleration structures that are driven by a relativistically intense laser at the moving plasma critical layer \cite{Sahai-PoP-2014}. These instabilities affect the energy-spectra of the accelerated ion-beams in propagating critical layer acceleration schemes \cite{Sahai-PRE-2013}\cite{Wilks-PRL-1992}. Specifically, using analytical theory and PIC simulations we look into three fundamental physical processes and their interplay that are crucial to the understanding of energy spectral control by making the laser-plasma ion accelerators stable. The interacting processes are (i) Doppler-shifted ponderomotive bunching \cite{Sahai-PoP-2014}\cite{Kaw-PoF-1971} (ii) potential quenching by beam-loading \cite{Sahai-PRE-2013} and (iii) two-stream instabilities. These phenomenon have been observed in simulations analyzing these acceleration processes \cite{Schlegel-PoP-2009}\cite{Bulanov-PoP-2012}\cite{Robinson-PPCF-2009}. From the preliminary models and results we present in this work, we can infer measures by which these instabilities can be controlled \cite{Sahai-IPAC-2012} for improving the energy-spread of the beams.

\end{abstract}

\maketitle

\section{Introduction}

In certain preliminary studies of laser-plasma interactions it has been shown that laser-plasma ion accelerators (LIA) hold the promise to source and accelerate light and heavy-ion beams with unique characteristics. The laser-plasma accelerated ion beams at moderately relativistic energies of up to a few GeVs are predicted to have properties of the order of $10^{10}$ particles in a beam dimension of the order of 100$\mu$m in each dimension at the source. Additionally these sources are much smaller than conventional accelerators of comparable energies, which also make them relatively inexpensive. As a result such source-accelerators may be affordably and widely applied to non-collider applications such as hadron tumor therapy, ultra-short ion-beam injectors, radiography, neutron production etc.

However, there are many major unsolved physics and technological challenges that need to be addressed in order to consider the various ion particle-streams from such sources as realistic beams for real-world applications. Among the major fundamental science challenges facing the LIAs a critical one is controlling the beam energy-spread, measured as the FWHM of the $N-\mathcal{E}$ vs $\mathcal{E}$ distribution ($\frac{\Delta\mathcal{E}_{beam}}{\mathcal{E}}$, the beam energy-spread relative to the energy at peak particle count N, in case the peak particle count is at a well-defined energy, in the accelerated $N-\mathcal{E}$ vs $\mathcal{E}$ distribution). Therefore schemes that use the propagating acceleration structures at the plasma critical layer that have been predicted to have a well-controlled velocity are under active research \cite{Sahai-PoP-2014}\cite{Sahai-PRE-2013}\cite{Wilks-PRL-1992}\cite{Schlegel-PoP-2009}\cite{Bulanov-PoP-2012}\cite{Robinson-PPCF-2009}\cite{Sahai-IPAC-2012}. Therefore it is crucial to understand the physical effects that affect the longitudinal phase-space of the beams accelerated off the propagating critical layer acceleration structures.

In this paper we present a preliminary analysis of the primary longitudinal instabilities of the acceleration structure driven at the moving critical layer during relativistically intense ($I_0 > 2.1 \times 10^{18}\frac{W}{cm^2}$) Ti:Sapphire ($\lambda_0\sim 800\mathrm{nm}$) laser-plasma interactions with overdense targets. A TEM laser mode propagating in vacuum at such intensities has transverse electric $E_{laser}=\sqrt{2 \eta I_0} = 2\pi\frac{m_ec^2}{e\lambda_0}a_0 \simeq 4 ~ \mathrm{TVm^{-1} (a_0=1)}$. At $\lambda_0\sim 800\mathrm{nm}$, the critical layer density is $n_{crit}=\frac{\pi m_e c^2}{e^2\lambda_0^2} \sim 1.7\times10^{21}\mathrm{cm^{-3}}$. These propagating critical layer ($\omega_{pe} = \omega_0$) acceleration structures are ideal for accelerating ion-beams \cite{Sahai-PoP-2014}\cite{Sahai-PRE-2013} for two reasons (a) the peak charge-separation acceleration field at the critical layer in the plasma, $E_{peak} \simeq 2\pi\frac{m_e c^2}{\lambda_0e}$ and the corresponding potential in the frame of the acceleration structure, $\Phi_{crit}$ is nearly as high as possible with Ti:Sapphire laser interacting with plasma (b) the laser-driven structures at the critical layer have controllable velocities that are essential for trapping the plasma-ions, as heavier particles require higher potential to get trapped, $eZ_{ion}\Phi_{crit} \geq (\gamma-1)M_{ion}c^2$\cite{Sahai-PoP-2014}.

The longitudinal instabilities of the acceleration structure that we analyze are as follows -

I. {\bf Doppler-shifted Ponderomotive Bunching} \cite{Sahai-PoP-2014}\cite{Sahai-PRE-2013}\cite{Kaw-PoF-1971} - The moving critical layer acceleration structures rely on relativistic laser pulse exciting the critical layer motion at a controlled velocity determined by the laser-plasma interaction parameters. In all such moving critical layer acceleration structures the laser pulse reflects off the {\em receding critical surface}, in the frame of the laser pulse the critical layer is propagating away from it. There are two physical effects due to this relative motion -  (i) the reflected laser-pulse frequency is red-shifted in the lab-frame, and (ii) the frequency and intensity of the laser in the frame of reference of the critical layer are relativistically modified. The effect (ii) also discussed in \cite{Sahai-PoP-2014} is beyond the scope of the current manuscript. 

The reflected laser pulse frequency is Doppler-shifted down from the laser pulse frequency to $\omega_{ref}$ and reflects with longitudinal phase velocity in opposite direction to the incident laser pulse. Under normal incidence, the {\em Doppler red-shifted reflection} from the moving critical layer, which interferes and mixes with the incoming laser pulse, creates a beat-modulation on the incoming laser pulse envelope, $a_{plasma}(x,t)$ at $\omega_0 - \omega_{ref}$. The beat-modulation of two counter-propagating pulses was intentionally induced and the resulting ponderomotive bunching used for the laser-plasma electron accelerators PBWA - Plasma Beat-Wave Accelerators \cite{PBWA}. This Doppler-shift beat modulation of the incoming laser pulse envelope leads to oscillations of the acceleration structure potential, $\Phi_{crit}(a_{plasma}(x,t))$ and its velocity, $v_{crit}(a_{plasma}(x,t))$. This effect is different from the ponderomotive bunching at stationary critical layer described in Ref.\cite{Kaw-PoF-1971} and occurs in addition to it. The characteristic parameter is the fraction of reflected laser amplitude relative to the incident laser amplitude, $\xi = a_{reflect} / a_{incident}$. As $\xi \rightarrow 1$ the effect becomes significant as the  resultant superposition amplitude during the destructive interference mixing node is nearly zero. The initial modulation of the acceleration structure velocity by $v_{crit}$ + $v_{doppler}$ feeds back to create a secondary Doppler-shift on the drive laser pulse. This secondary Doppler-shift causes a secondary acceleration structure velocity modulation which again feeds back onto the Doppler-shift further modulation of laser-envelope and the velocity resulting in a continual feedback. This continual feedback is more accurately modeled with PIC codes modeling laser-plasma interactions at the critical layer. The instability of the acceleration structure leads to an instability of the energy-spectra of the beam accelerated off this acceleration potential. It may be possible to control this instability by using oblique angles or frequency-chirp \cite{Sahai-IPAC-2012} so that the reflected pulse has minimal mixing with the incident laser pulse.

\begin{figure*}
	\begin{center}
   	\includegraphics[width=6.5in]{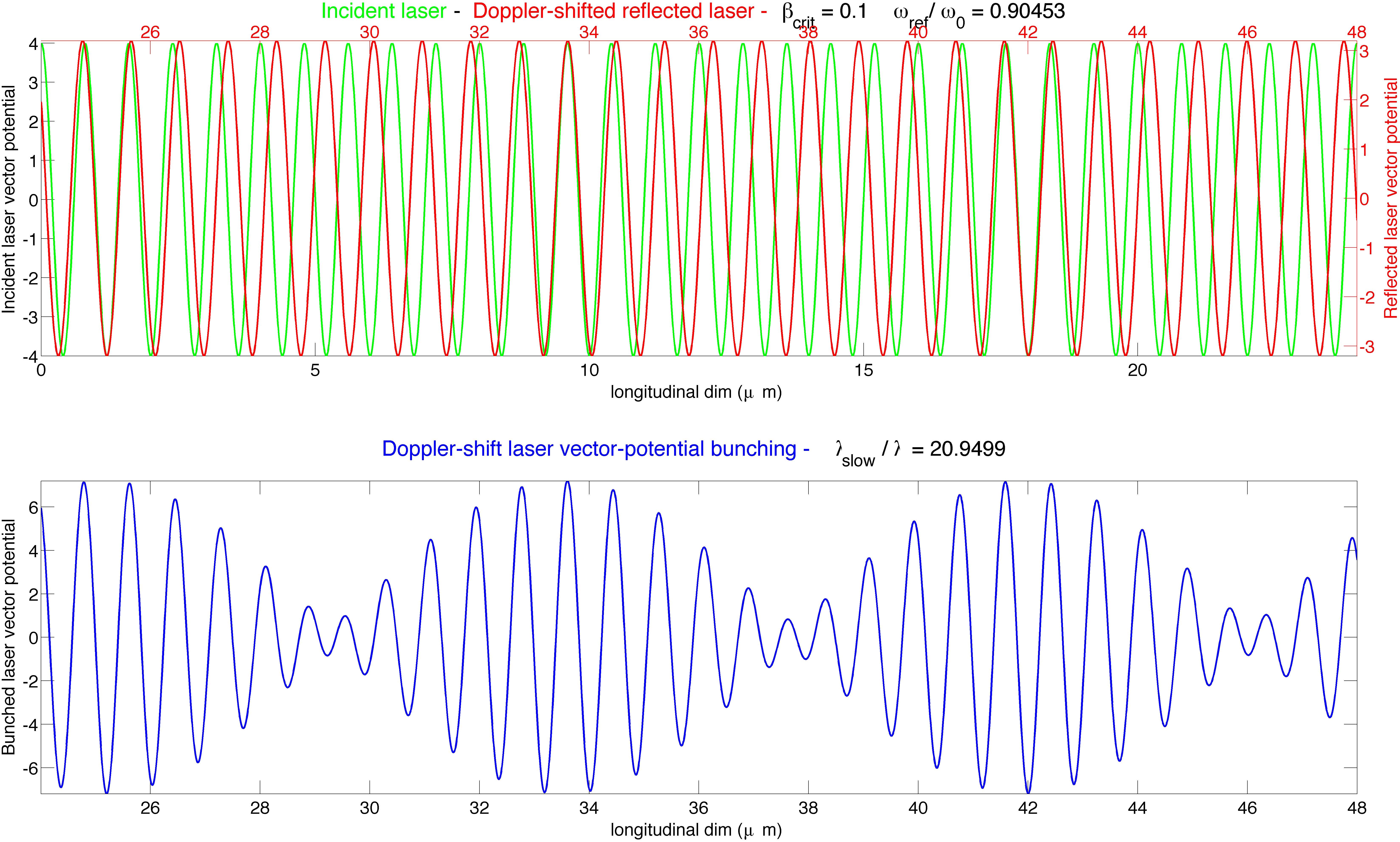}
	\end{center}
\caption{ {\it Doppler-shifted Ponderomotive bunching from the analytical expression}. Doppler-shifted ponderomotive bunching with $\xi = 0.8$ and $a_0 = 4.0$, {\color{green} GREEN} curve is $a_{inc}(t)$ the incident laser vector potential, {\color{red} RED} is $a_{ref}(t)$ the Doppler shifted reflected laser vector, {\color{blue} BLUE} is the bunched laser vector potential upon superposition.}
\label{fig:analytic-DSPB}
\end{figure*}

II. {\bf Acceleration potential quenching by uncontrolled beam-loading} - This effect occurs due to saturation of beam-loading by the accelerated beam, $\mathcal{E}_{beam} \simeq \mathcal{E}_{\Phi}$. In most of the moving critical layer schemes the number of particles being loaded on the charge-separation potential is uncontrolled \cite{Schlegel-PoP-2009}. The acceleration potential initially builds up as the charge separation is driven at the critical layer by the laser. When the potential reaches the trapping and reflection threshold it transfers energy to the trapped plasma-particle beam and correspondingly the potential energy in charge-separation is depleted. The characteristic parameter is the ratio of total energy in the accelerated beam to the energy in the acceleration potential at the trapping threshold, which can be denoted by $\kappa = \mathcal{E}_{beam} / \mathcal{E}_{Th}$. Therefore, $\kappa = $ (total ion-beam energy) / (energy in electron charge-separation potential $>$ trapping threshold). As $\kappa \rightarrow 0.1$, approaches 10 percent this effect becomes significant and energy transfer significantly depletes the potential, reducing it to below the threshold. Once the potential is below threshold the trapping of plasma-particles stops. The trapping resumes again when the potential is above the threshold that may take a significant fraction of time, assuming that the laser pulse is still driving the critical layer and is itself stable. As a result the total number of particles reflecting off the acceleration potential undergo an oscillations depending upon the factor $\kappa$. It may be possible to control this instability by controlling the concentration of the accelerated species that beam-loads the acceleration potential.

III. {\bf Two-stream instabilities} - we may identify two different types of two-stream interactions in the moving critical layer laser-plasma acceleration processes, (a) Buneman-like electron-ion instability and (b) ion-ion counter- streaming beam instability between ion-beam and plasma-ions \cite{Bulanov-PoP-2012}. This instability arising due to velocity difference between the coupled species leads to a longitudinal density and velocity bunching of the plasma particles. The characteristic parameter of this instability is the growth-rate $\gamma_{ts}$ and the frequency of density perturbations, $\omega_{ts}$. The parameters are dependent upon the density ratio between counter-streaming particle beams and the mass ratio.

(a) {\it Buneman-like instability} - A fraction of fast-electrons driven by the longitudinal laser ponderomotive force that escape the acceleration structure charge-separation potential and propagate longitudinally away from it, interact with the target plasma ahead of the critical layer. The coupling of energy from the ponderomotive fast electrons to the background ions can significantly modulate the energy of the plasma ions. When pre- modulated plasma-ions are trapped and reflected, the accelerated beam has an initial energy modulation.

(b) {\it Ion-ion counter-streaming instability} - The ion-beam accelerated off the moving acceleration structure propagates in plasma with relatively stationary plasma ions. The relative velocity between the ion beam and the background plasma-ions leads to the development of a counter-streaming instability.

\begin{figure*}
	\begin{center}
   	\includegraphics[width=6.5in]{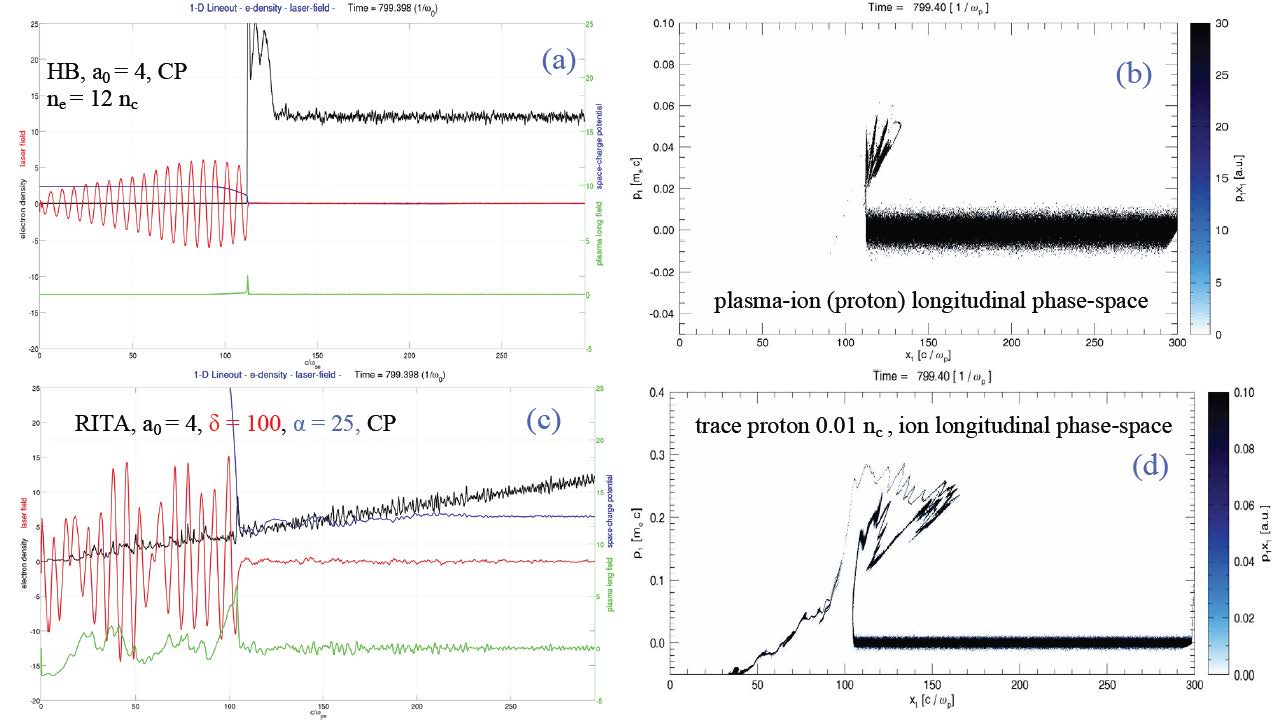}
	\end{center}
\caption{ {\it Doppler-shifted Ponderomotive bunching from PIC simulations of HB and RITA}. PIC simulations of HB and RITA showing Doppler-shifted ponderomotive bunching in the laser-field and in the accelerated beam spectrum, the laser vector potential is in {\color{red}RED}, the plasma field is in {\color{green}GREEN}, electron density is in BLACK.}
\label{fig:PIC-DSPB-HB-RITA}
\end{figure*}

\section{Doppler-shifted Ponderomotive bunching}
The laser pulse incident at the critical layer characterized by its vector potential in 1D is written as a product of the slowly-varying envelope profile $a_{env}(x,t)$ and the fast oscillations $cos(\omega_0 t + k_0x)$ at the laser frequency, $(\omega_0, k_0)$, $a_{inc}(x,t) = a(x,t) cos(\omega_0 t + k_0x)$. The laser pulse reflected off the critical layer is then, $a_{ref}(x,t) = \xi(t) a_{env}(x,t) cos(\omega_{ref}(t) t - k_{ref} x)$. The interference between the incident and the reflected wave leads to a beat-pattern in the laser-pulse envelope. This beat-pattern in the envelope leads to the {\it bunching of the ponderomotive force} driving the charge-separation. The {\it Doppler-shifted ponderomotive bunching} effect is relevant if the pulse length is longer than half-wavelength of the beat-wave, $\tau_p > \frac{1}{2} 2\pi (\omega_0-\omega_{ref})^{-1}$.

$a_{bunched}(x,t) = a_{inc}(x,t) + a_{ref}(x,t)$ \\= $a_{inc}(x,t)$ \\ $\times Re\left( exp[ j(\omega_0 t + k_0 x)] + \xi(t) exp [j(\omega_{ref}(t) t - k_{ref}(t) x)] \right)$ 

For comparison to the PIC simulations in the spatial-domain, when $\xi(t) \simeq 1$ (reflection coefficient), we can rewrite the superposition, as, 
\begin{eqnarray}
&\nonumber a_{bunched}(t) \simeq 2 a_{env}(x,t) \\
& \times cos\left(\frac{\omega_0 - \omega_{ref}(t)}{2} t + \frac{k_0 - k_{ref}(t)}{2} x\right) cos(\omega_0 t + k_0 x)
\label{bunched-exp}
\end{eqnarray}

$\omega_{ref}(t), k_{ref}(t)$ = Doppler-shifted reflected frequency and wave-vector off the receding critical layer. 
$\omega_{ref}(t) = \frac{\omega_0}{\gamma_{crit} (1 + \beta_{crit})}$, 
$\beta_{crit}(t) = v_{crit}(t) / c$,
$\gamma_{crit} = (1 - \beta_{crit}^2)^{-1/2}$,
$k_{ref}(t) = \frac{k_0}{\gamma_{crit} (1 - \beta_{crit})}$\\

The longitudinal ponderomotive force is $F_p \propto a_{env}^2$, therefore the ponderomitve force is bunched due to the Doppler-shifted reflection from a moving critical layer at a frequency of $\omega_{ref}-\omega_0$. So, a flat-top laser would have a varying ponderomitve force because of the effect of {\it Doppler-shifted ponderomotive bunching}.

Since $(\omega_{ref}-\omega_0) \propto \beta_{crit}$, at higher velocity this effect applies to shorter pulses as the bunching occurs at shorter time and spatial scales. 

If the critical-layer velocity is not constant then the reflected wave parameters have the same time-dependence as the critical layer velocity. The time-variation of the velocity is important under self-consistency and gives rise to a feedback on the Doppler-shifting of the frequency further modulating the velocity and degrading the ion spectra accelerated using the propagating acceleration structures.

The superposition analytical expression in eq.\ref{bunched-exp} is simulated as shown in fig.\ref{fig:analytic-DSPB}  where the resultant laser-potential is beat-wave modulated with a slow envelope. At $\beta_{crit} = 0.1$ and $\xi = 0.8$ the reflected laser wave-vector is $k_{ref}  = 0.9k_0$, $\frac{(k_0 - k_{ref})}{2} \simeq k_0 / 21$. In the beat-wave, there are $\sim 21 \lambda_0$ cycles within a full-cycle and $\sim 10.5 \lambda_0$ within a single beat cycle. Thus analytical model shows that a flat-top laser pulse envelope is modified due to the effect of beat-wave. \\

The effect of Doppler-shifted ponderomotive-bunching is shown using 1-D PIC simulations (parameters in figure) of Hole-Boring (HB) \cite{Wilks-PRL-1992}\cite{Schlegel-PoP-2009}\cite{Bulanov-PoP-2012}\cite{Robinson-PPCF-2009}, Fig.\ref{fig:PIC-DSPB-HB-RITA}(a),(b) and RITA \cite{Sahai-PRE-2013}\cite{Sahai-PoP-2014} schemes Fig.\ref{fig:PIC-DSPB-HB-RITA}(c),(d). In Fig.\ref{fig:PIC-DSPB-HB-RITA}(a),(c) from PIC simulation snapshots we see that the resultant laser field at the critical layer is modulated once the reflection interferes with the incident laser in both the moving critical layer schemes. The interference between the incident and the reflected wave creates a beat-wave that has the slowly-varying envelope. The modulation of the velocity and the acceleration potential can be seen in the proton longitudinal phase-space in HB Fig.\ref{fig:PIC-DSPB-HB-RITA}(b),(d). It should be noted that the modulated velocity due to beat-wave leads to another Doppler-shift which further modulates the laser-pulse, resulting in a feedback loop on the envelope of the laser pulse resulting in an instability and as a result the acceleration structure velocity and potential are prone to instability. In PIC simulations we observe that with {\it frequency chirped-pulse} the Doppler-shifted envelope modulation and ponderomotive-bunching may be controlled \cite{Sahai-IPAC-2012}.

\begin{figure*}
	\begin{center}
   	\includegraphics[width=6.5in]{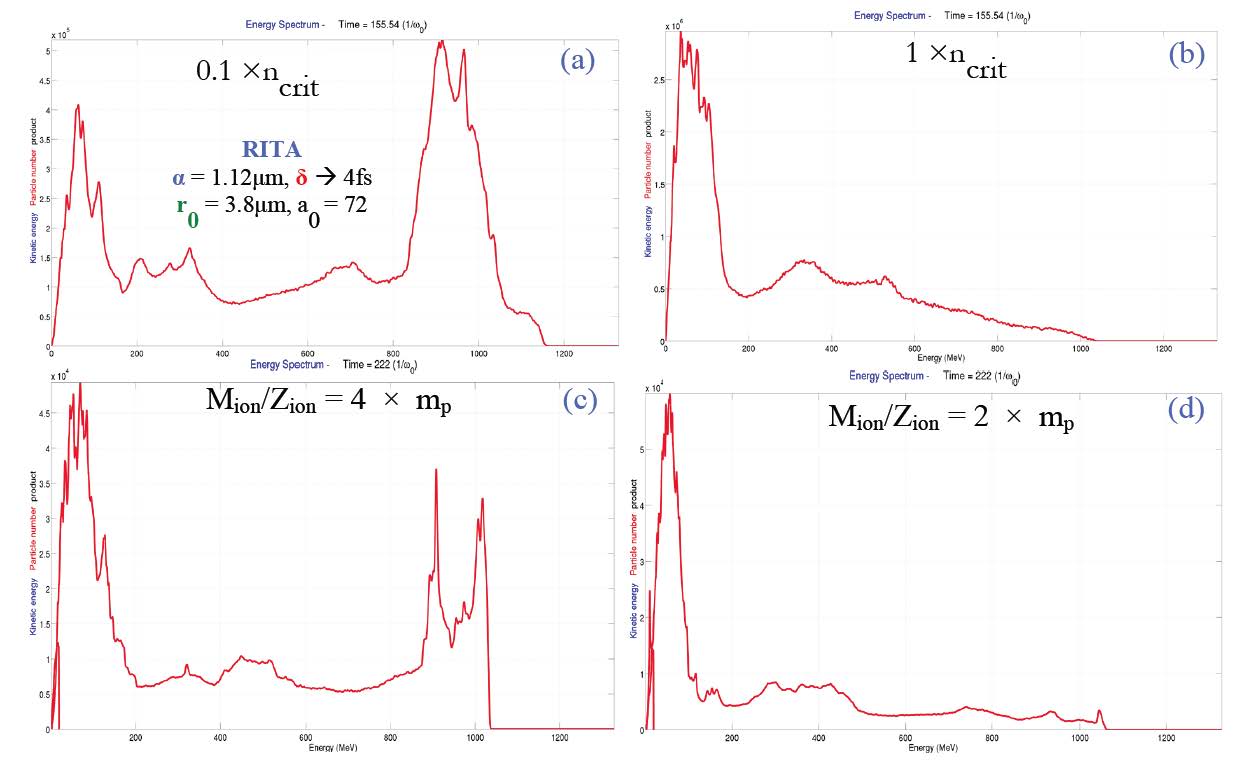}
	\end{center}
\caption{ {\it Beam-loading PIC simulations of RITA}. Top-row (a),(b) is with {\it IMMOBILE BACKGROUND IONS} shows the effect of accelerated species concentration (trace-proton) changed from 0.1 to 1.0$n_c$. Bottom-row (c),(d) with {\it REALISTIC-MASS BACKGROUND IONS} demonstrates the effect of unwanted beam- loading by the background-ions that lowers the space-charge potential when the background ions mass is changed from 4 to 2 $m_p$.}
\label{fig:PIC-BL-RITA}
\end{figure*}

%
\section{Potential quenching by excessive beam-loading}

The effect of the beam-loading by the accelerated particles on the acceleration structure potential is demonstrated using 2D PIC simulations of RITA in Fig.\ref{fig:PIC-BL-RITA} (parameters in figure) for $M_{ion}/Z_{ion}=m_p$ (proton) beam accelerated to around a GeV. The beam-loading leads to the quenching of the acceleration potential of the acceleration structure.\\

(a)	{\it Excessive beam-loading -} is shown in Fig.\ref{fig:PIC-BL-RITA}(a),(b) by varying the trace-proton density from 0.1 $n_c$ to 1.0 $n_c$. The background ions are stationary. It is clear that under heavy loading the acceleration-time is smaller. Under heavier loading of the acceleration structure potential (b) fewer particles are accelerated, disrupting the significant mono-energetic feature in the spectra, (a).\\

(b)	{\it Undesired beam-loading -} is shown in Fig.\ref{fig:PIC-BL-RITA}(c),(d), potential traps and couples to ions of realistic $M_{ion}/Z_{ion}$ ratio. When the mass-to-charge ratio is 4$m_p$, the potential is not high enough to significantly perturb the background ions whereas when it is 2$m_p$ the structure's potential energy is partitioned into trace-protons and background ions.

\begin{figure*}
	\begin{center}
   	\includegraphics[width=6.5in]{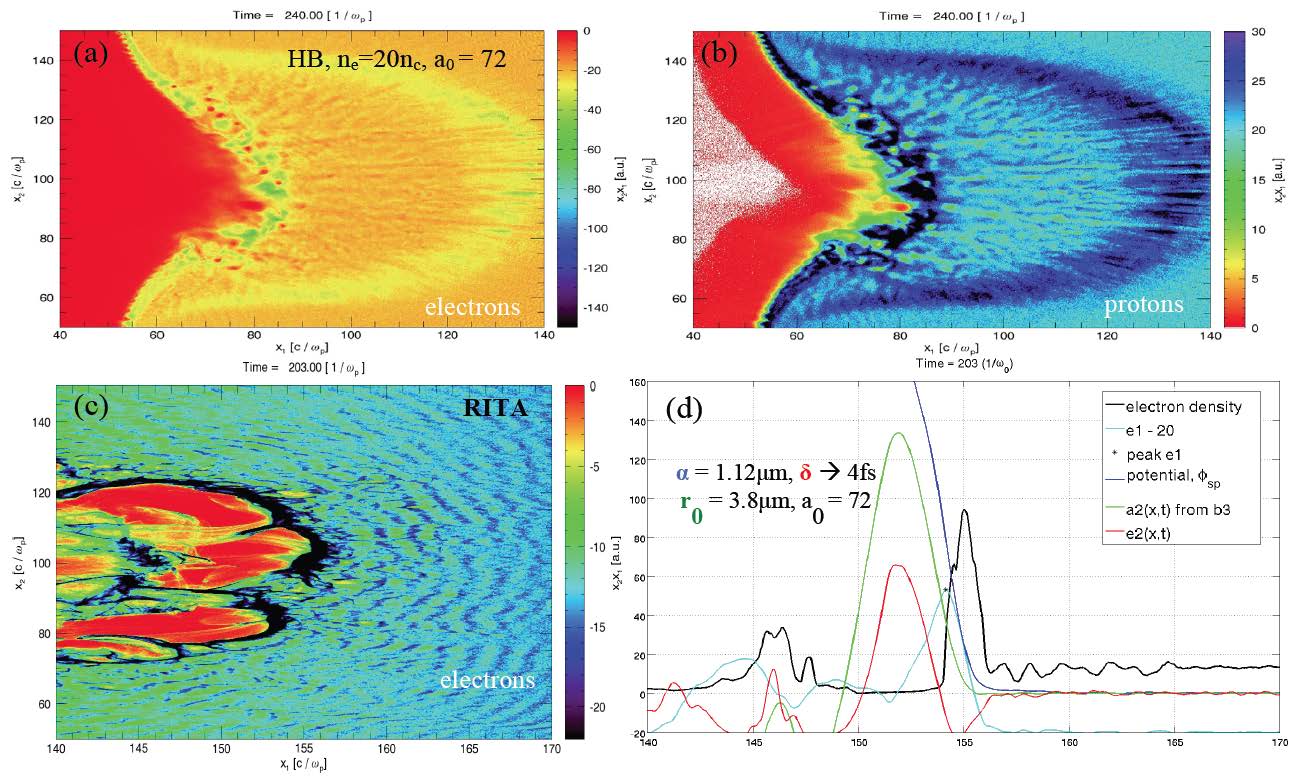}
	\end{center}
\caption{ {\it Pre-modulation of accelerated species by fast-electrons}. Top-row is real-space density of Hole-Boring (a) electrons and (b) protons showing the pre-modulation of the ions much ahead of the hole-boring front. Bottom-row is real-space density of RITA (c) electrons and (d) lineout of the on-axis electron density.}
\label{fig:PIC-TS-HB-RITA}
\end{figure*}

\section{Pre-Modulation by Fast electrons - 2-Stream Instabilites}

We show the effect of the ponderomotively driven fast plasma electrons that break-away from the potential of the acceleration structure. These fast electrons interact with the plasma ions much ahead of the laser-driven structure and excite a two-stream instability and pre-modulate them, well before the structure reaches these ions. This can be seen in the case of HB in Fig.\ref{fig:PIC-TS-HB-RITA} (a),(b) and RITA in Fig.\ref{fig:PIC-TS-HB-RITA}(c),(d). From these snapshots in Fig.\ref{fig:PIC-TS-HB-RITA} it is seen that the propagating electron beam-lets excite ion density modulations. Therefore, when the acceleration structure traps and reflects the pre-modulated ions, the ions accelerated have an initial spread. 

\section{Discussion}
We have shown by modeling and PIC simulations, some of the significant longitudinal laser-plasma interaction instabilities affecting the laser-plasma ion acceleration structures created at moving critical layer. These longitudinal instabilities thereby disrupt the energy spectra of the accelerated species and hinder the mono-energetic acceleration of ions in laser-plasma accelerators. In future we will further explore the nature of these dynamical effects and the instabilities.

\section{Acknowledgement}

Work supported by National Science Foundation, NSF-PHY-0936278 and Department of Energy, DE-SC-0010012. We acknowledge the use of 256-node {\it Chanakya} cluster of the Pratt school of engineering at Duke University.


\end{document}